\documentclass[openacc]{rstransa}
\usepackage{graphicx,url,twoopt}
\usepackage[varg]{txfonts}           
\usepackage{hyperref}
\usepackage{booktabs}
\usepackage{url}            
\usepackage{pdfcomment}              
\usepackage{acronym} 
\usepackage{epstopdf}                
\usepackage{bm}
\usepackage{longtable}
\usepackage{caption}
\usepackage{mathtools}
\usepackage{esvect}
\usepackage{amsmath}
\usepackage{multirow}
\usepackage{adjustbox}
\usepackage{amssymb} 
\usepackage{amsfonts} 
\usepackage{amsthm} 
\usepackage{endnotes} 
\usepackage{setspace}
\usepackage{verbatim} 
\usepackage{times} 
\usepackage{helvet} 
\usepackage{courier} 
\usepackage{wrapfig} 
\usepackage{dcolumn}

\newcommand{\LT}{$L_{\text{X}}-T_{\text{X}}$}
\newcommand{\LY}{$L_{\text{X}}-Y_{\text{SZ}}$}
\newcommand{\LMgas}{$L_{\text{X}}-M_{\text{gas}}$}
\newcommand{\MgasT}{$M_{\text{gas}}-T_{\text{X}}$}
\newcommand{\YT}{$Y_{\text{SZ}}-T_{\text{X}}$}
\newcommand{\RT}{$R_{50\%}-T_{\text{X}}$}
\newcommand{\RMgas}{$R_{50\%}-M_{\text{gas}}$}
\newcommand{\Lx}{$L_{\text{X}}$}
\newcommand{\Ysz}{$Y_{\text{SZ}}$}

\newcommand{\Lbcg}{$L_{\text{BCG}}$}
\newcommand{\Tx}{$T_{\text{X}}$}

\titlehead{Review}
\jname{rsta}
\Journal{Phil. Trans. R. Soc}

\begin{document}
\title{Galaxy clusters as probes of cosmic isotropy}

\author{
Konstantinos Migkas$^{1,2}$}
\address{$^{1}$Leiden Observatory, Leiden University, PO Box 9513, 2300 RA Leiden, the Netherlands\\
$^{2}$SRON Netherlands Institute for Space Research, Niels Bohrweg 4, NL-2333 CA Leiden, the Netherlands}
\subject{cosmology, astrophysics}

\corres{Konstantinos Migkas\\
\email{kmigkas@strw.leidenuniv.nl}}

\begin{abstract}
Scaling relations of galaxy clusters are a powerful probe of cosmic isotropy in the late Universe. Owing to their strong cosmological dependence, galaxy cluster scaling relations can obtain tight constraints on the spatial variation of cosmological parameters, such as the Hubble constant ($H_0$), and detect large-scale bulk flow motions. Such tests are crucial to scrutinise the validity of $\Lambda$CDM in the local Universe and determine at what cosmic scales (if any) extra-galactic objects converge to isotropy within the Cosmic Microwave Background rest frame. This review describes the methodology for conducting cosmic isotropy tests with cluster scaling relations and examines possible systematic biases. We also discuss the results of past studies that reported statistically significant observed anisotropies in the local Universe. Finally, we explore the future potential of cluster scaling relations as a cosmic isotropy probe given the large amount of multi-wavelength cluster data expected in the near future.
\end{abstract}


\begin{fmtext}
\section{Introduction}
The Cosmological Principle (CP) postulates that the Universe is statistically isotropic and homogeneous over sufficiently large scales. The primary experimental evidence for this assumption comes from Cosmic Microwave Background (CMB) observations \cite{planck18}. CMB data support the existence of a rest frame (the so-called CMB rest frame) in which cosmic radiation in the very early Universe was highly isotropic. To reach this rest frame, one needs to remove the main anisotropic feature of the CMB; the well-known, dominant CMB dipole. The latter is presumed to arise due to our peculiar motion as observers within the CMB rest frame and the Doppler effect. Adopting the fully kinematic nature of the CMB dipole, one concludes our Solar system moves with $370$ km/s towards the Galactic coordinates $(l,b)\approx (264^{\circ}, +48^{\circ})$ within the CMB rest frame \cite{planck20}. 
\end{fmtext}

\maketitle

The adoption of the CP in observational cosmology implies that cosmic matter in the late, low$-z$ Universe should share the same rest frame as radiation in the early, high$-z$ Universe. In other words, extra-galactic objects, such as galaxies, galaxy clusters, and supernovae (SN), should converge to statistical isotropy within the CMB rest frame when large cosmic volumes are considered. However, the CP alone does not define the scale at which convergence to isotropy should be achieved. For this, one needs individual cosmological models that predict the level of apparent anisotropy as a function of scale, as different models return different predictions. Hence, when one scrutinises the apparent (an)isotropy of the local Universe, one essentially tests the validity of specific cosmological models, and not the CP itself.

If a cosmological model is intrinsically isotropic, apparent anisotropies might emerge from unaccounted bulk flow motions.\footnote{The average peculiar velocity of a given cosmic volume, estimated by the peculiar velocities of individual extra-galactic objects within this volume.} Such bulk flows can bias the inference of the Hubble flow $z_{\text{Hubble}}$ (the notation $z\equiv z_{\text{Hubble}}$ will be used hereafter) from the measured $z_{\text{obs}}$ if the true peculiar velocities are not accurately known or not accounted for. Subsequently, the inferred cosmological distance from $z$ will also be biased. The properties of bulk flows strongly depend on the degree of matter homogeneity at each cosmic scale. Consequently, the bulk flow amplitude as a function of scale varies for different models. 

The standard cosmological model, $\Lambda$CDM, predicts that at $\gtrsim 200$ Mpc scales bulk flows should dissipate to $\lesssim 140$ km/s (e.g., \cite{watkins23}). If such bulk flows are ignored, they can result in an observed $\lesssim 3\%$ anisotropy in the redshift-distance relation and the Hubble constant ($H_0$) within the CMB rest frame. Subsequently, larger apparent $H_0$ anisotropies or bulk flows at $\gtrsim 200$ Mpc scales would pose a challenge for $\Lambda$CDM. Interestingly, several studies have reported such observed anisotropies at $z\lesssim 0.06-0.2$, providing indications that the matter rest frame might not converge to the CMB rest frame at $\gtrsim 200$ Mpc (e.g., \cite{colin11, carrick,colin19,salehi,conville,cowell, sorrenti, watkins23,whitford,hoffman, hu24,yarahmadi}). On the other hand, only a few of these findings exhibit sufficiently high statistical significance, while most are at only mild tension ($\lesssim 3\sigma$) with $\Lambda$CDM. As a result, the consistency of the local Universe kinematics with $\Lambda$CDM remains rather ambiguous and a topic of ongoing debate. Thus, it is crucial to develop novel, powerful techniques to scrutinise the assumption of isotropy in the late Universe and conclusively test the validity of $\Lambda$CDM at local scales.

\section{Galaxy clusters as probes of local Universe isotropy}
Galaxy clusters are one of the best probes to trace the isotropy of the redshift-distance relation at $\sim 200-2000$ Mpc ($z\sim 0.05-0.5$) scales. Clusters are uniformly distributed across the entire extra-galactic sky while we have high-quality data for thousands of clusters at these cosmic scales. In contrast, full-sky galaxy surveys are typically restricted to $z\lesssim 0.1$ scales (e.g., \cite{tully}). Moreover, although SNIa samples (e.g., \cite{scolnic}) cover a wide range of distances ($z\sim 0.01-1.5$), they suffer from strongly inhomogeneous sky distributions that can bias isotropy studies (e.g., \cite{hu24}). As a result, galaxy clusters currently offer the best option for studying bulk flows and cosmological anisotropies at $z\sim 0.05-0.5$ scales, where, according to $\Lambda$CDM, extra-galactic objects should behave isotropically within the CMB rest frame.

\subsection{Cosmology-dependent cluster properties}
A key element of galaxy clusters is that they are observed across most of the electromagnetic spectrum. Thus, numerous multi-wavelength cluster quantities can be determined independently. For instance, in X-rays, the hot ionised cluster gas is detected by its bremsstrahlung continuum and line emission. This allows one to measure the cluster's X-ray luminosity (\Lx), gas mass ($M_{\text{gas}}$), and X-ray isophotal radius ($R_{50\%}$). In optical and infrared frequencies, one can examine the galaxy population of clusters and estimate the luminosity of the brightest central galaxy (\Lbcg). Employing optical weak lensing data, the total cluster mass ($M_{\text{tot}}$) can also be estimated. In microwave frequencies, the total integrated Compton parameter (\Ysz, equivalent to the total cluster gas pressure) is measured using the thermal Sunyaev-Zeldovich effect. The latter is caused by the inverse Compton scattering of CMB photons as they travel through the hot cluster gas. 

The above-mentioned cluster properties can only be estimated with knowledge of the cluster's distance. However, an absolute calibration of cluster distances has yet to be available. Hence, assuming a $\Lambda$CDM Universe, one can infer the cluster's distance from its $z$ as 
\begin{equation}
D_L(z) = \dfrac{c(1+z)}{H_0} \int_0^{z} \frac{dz'}{E(z')}\quad \text{and}\quad D_A(z)=\frac{D_L(z)}{(1+z)^{2}},
\end{equation}
where $D_L(z)$ and $D_A(z)$ are the luminosity distance and angular diameter distance respectively, $c$ is the speed of light, $E(z)=\sqrt{\Omega_{\text{m}}(1+z)^3+\Omega_{\Lambda}}$ is the normalised Hubble parameter, and $\Omega_{\text{m}}$ and $\Omega_{\Lambda}$ are the normalised matter and dark energy densities respectively. As a result, the inference of these galaxy cluster properties from observables strongly depends on the fiducial $H_0$ value, making such cluster properties \emph{cosmology-dependent}. Moreover, the inference of $z$, and consequently $D_L(z)$ and $D_A(z)$, depends on the observed $z_{\text{obs}}$ and the peculiar velocity $z_{\text{pec}}$ as:
\begin{equation}
1+z_{\text{obs}}=(1+z)(1+z_{\text{pec}}).
\label{redshifts}
\end{equation}
Thus, bulk flows can bias the cluster distance if the former are ignored, i.e., it is assumed that $z_{\text{obs}}\approx z$. This bias then propagates to all cosmology-dependent cluster properties.

\subsection{Cosmology-independent cluster properties}
The key cluster property to test cosmic isotropy is the gas temperature ($T_{\text{X}}$). The latter is observationally determined through X-ray spectroscopy and its measurement does not depend on cosmological parameters or peculiar velocities.\footnote{The only cosmological dependency of the \Tx\ measurement comes from the selection of the cluster spectra extraction region which usually depends on first estimating the radius of the cluster. However, the final \Tx\ dependency on $H_0$ is $T_{\text{X}}\sim H_0^{-0.07}$ (see Appendix B in \cite{migkas21}). The \Tx\ dependency on bulk flows is similarly weak (see Appendix E.4 in \cite{migkas21}).} Therefore, \Tx\ is \emph{cosmology-independent}. \Tx\ is a measure of the kinetic energy of the gas and reflects the depth of the cluster's gravitational well. The latter is also traced by the kinetic energy of the galaxies, which is characterised by the galaxy velocity dispersion $\sigma_{\text{vel}}$, measured in optical/infrared wavelengths. $\sigma_{\text{vel}}$ is also determined without any cosmological assumptions and can be an alternative to \Tx\ as a cosmology-independent cluster property to be used for testing cosmic isotropy. Finally, the optical richness of galaxy clusters ($\lambda$) is another useful cosmology-independent quantity for isotropy tests.

\subsection{Galaxy cluster scaling relations}
All the multi-wavelength cluster properties mentioned above scale with the total cluster mass. As a result, they scale with each other, giving rise to the so-called \emph{galaxy cluster scaling relations}. Most such relations are theoretically motivated (see e.g., \cite{kaiser}) and have been established observationally for at least two decades (see \cite{giodini,lovisari22} for a review on galaxy cluster scaling relations). They are described by single power laws that in logarithmic space take the form:
\begin{equation}
    \log{Y}+ C_{YX}\log{E(z)}=A_{YX}+B_{YX}\log{X}\pm \sigma_{\text{intr}, YX}
    \label{basic_scaling}
\end{equation}
where $Y$ and $X$ are the two scaling cluster properties, $A_{YX}$, $B_{YX}$, and $\sigma_{\text{intr}, YX}$ are respectively the intercept, slope, and intrinsic scatter of the $Y-X$ scaling relation, and $C_{YX}$, together with the $\log{E(z)}$ term, account for the redshift evolution of the relation. When one uses a cosmology-dependent and cosmology-independent cluster property as $Y$ and $X$, the cosmological dependence of such a scaling relation maximises. 
\begin{figure}[h]
\centering \includegraphics[width=3.3in]{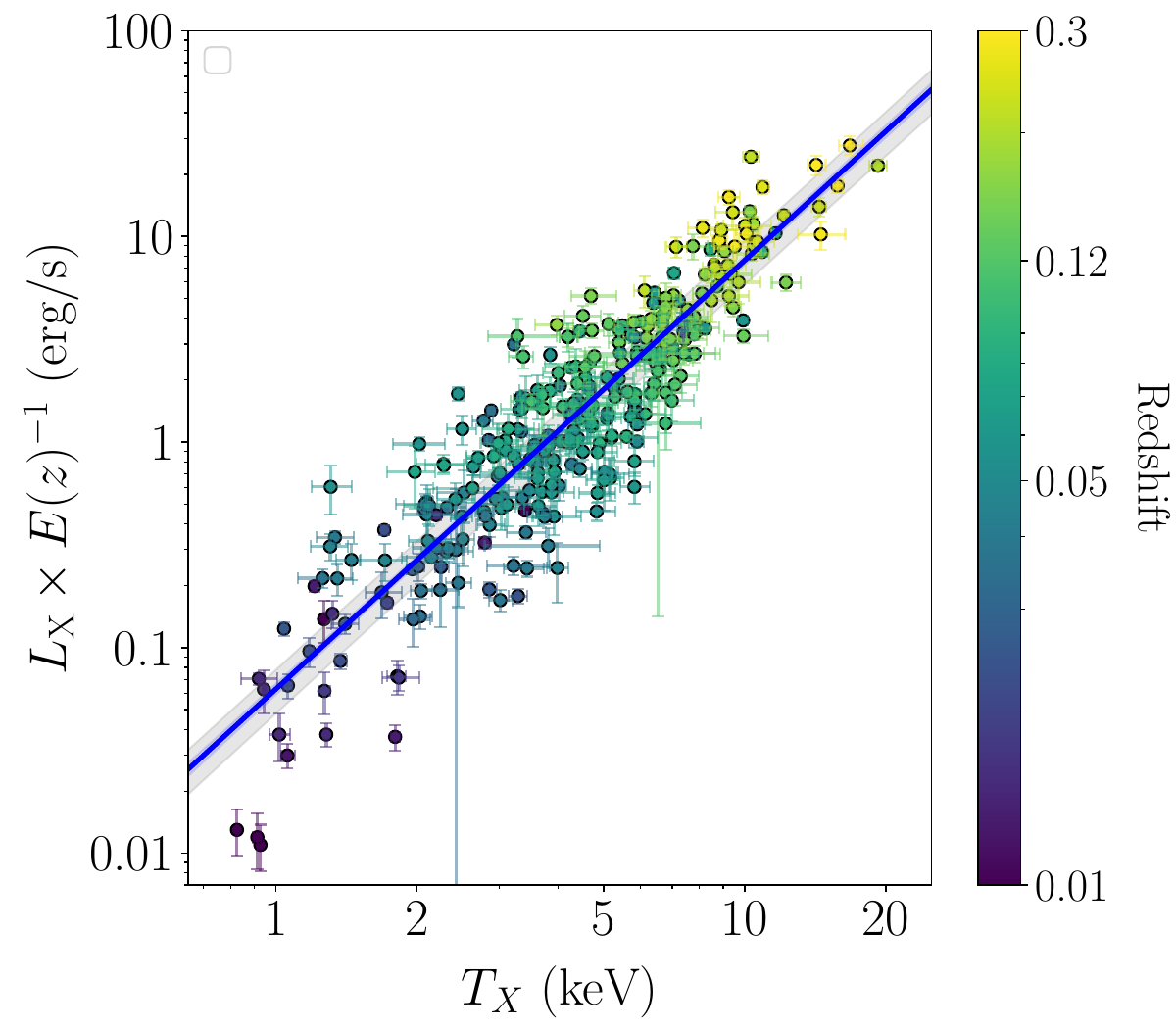}
\caption{\LT\ scaling relation for the 313 clusters presented in \cite{migkas20} and \cite{migkas21}. The colour of data points corresponds to the cluster redshift. The blue and grey shaded areas represent the statistical and total scatter, respectively.}
\label{L-T-plot}
\end{figure}
For example, let us consider the \LT\ relation (Fig. \ref{L-T-plot}). We know that $L_{\text{X}}=4\pi D_L(z)\times K_{\text{corr}}f_{\text{X}}$, where $f_{\text{X}}$ is the observed X-ray cluster flux and $K_{\text{corr}}$ is the K-correction that converts the \Lx\ from the observer's rest frame to the cluster's rest frame. Therefore, Eq. \ref{basic_scaling} can be written as
\begin{equation}\label{decomposed_LT}
\log{f_{\text{X}}}+ 2\log{D_L(z)} + C_{LT}\log{E(z)}+\log(4\pi K_{\text{corr}}) =A_{LT}+B_{LT}\log{T_{\text{X}}}\pm \sigma_{\text{intr}, LT}.
\end{equation}
All terms that include $E(z)$ and numerical constants can be summarised into a single term $G(\Omega_{\text{m}},C_{LT},z)$.\footnote{$G(\Omega_{\text{m}},C_{LT},z)=2\log{\left[c(1+z)\int_0^{z} \frac{dz'}{E(z')}\right]} + C_{LT}\log{E(z)}+\log(4\pi K_{\text{corr}})$} Eq. \ref{decomposed_LT} can be then rewritten separating the cosmology-dependent and cosmology-independent terms:
\begin{equation}
    2\log{H_0}+A_{LT} + G(\Omega_{\text{m}},C_{LT},z) = \log{f_{\text{X}}}-B_{LT}\log{T_{\text{X}}} \pm \sigma_{\text{intr}, LT}.
    \label{LT-anisotropy}
\end{equation}
On the right side of Eq. \ref{LT-anisotropy}, $f_{\text{X}}$ and \Tx\ are observables and directly measured quantities. On the contrary, $B_{LT}$ and $\sigma_{\text{intr}, LT}$ are nearly completely insensitive to cosmological changes. On the left side of Eq. \ref{LT-anisotropy}, there is strong cosmological dependence due to the $2\log{H_0}$ and $G(\Omega_{\text{m}},C_{LT},z)$ terms. The intercept $A_{LT}$ is not a cosmological quantity but it is completely degenerated with $H_0$ since $A_{LT}$ is not known independently of cosmological assumptions. 
\subsection{Testing cosmological parameter isotropy and bulk flows}
In an isotropic Universe, the best-fit $2\log{H_0}+A_{LT} + G(\Omega_{\text{m}},C_{LT},z)$ sum should not spatially vary. One can scrutinise the isotropy hypothesis and investigate the directionality of cosmological parameters (e.g., $H_0$) and bulk flows by constraining Eq. \ref{LT-anisotropy} in independent sky patches. To do so, sufficiently large and uniform cluster samples should be employed. Such an approach reveals how much each sky direction deviates from the average, all-sky behaviour. This deviation can then be translated to the necessary variation of cosmological parameters to reconcile each direction with the rest of the sky. This approach does not assume the anisotropy shape. Alternatively, one can fit Eq. \ref{LT-anisotropy} to the full, all-sky cluster sample while introducing an anisotropic component with an assumed shape in the scaling relation model, e.g., a dipole anisotropy. In this case, the extra free parameters are the direction and amplitude of the anisotropy. Each approach can be applied to different $z$ bins to characterise any retrieved anisotropy as a function of cosmic scale. As explained in Sect. \ref{H0_anis}, \ref{bulk flows}, and \ref{Omega}, using different $z$ bins one can eventually distinguish $H_0$ and $\Omega_{\text{m}}$ anisotropies, as well as bulk flows since these cosmological phenomena impact the observed anisotropy of Eq. \ref{LT-anisotropy} differently. Thus, the degeneracy between the individual left-side terms of Eq. \ref{LT-anisotropy} can be broken and one can draw conclusions on individual cosmological parameters.

Every cluster scaling relation can be decomposed to observables and cosmological parameters as in Eq. \ref{LT-anisotropy}. Each time, the cosmological sensitivity of the scaling relation changes according to how the utilised cluster properties depend on cosmic distance. For example, $M_{\text{gas}}\propto D_A^{\frac{5}{2}}$; as a result, the \MgasT\ relation leads to the strongest $H_0$ dependence with $2.5\log{H_0}$.\footnote{The dependence on $\Omega_{\text{m}}$ and other parameters through $G$ is more complicated and also depends on the redshift evolution of the relation as explained in Sect. \ref{Omega}} Similarly, the \RT\ relation leads to a $\log{H_0}$ dependence. Furthermore, one can study scaling relations between two cosmology-dependent properties and still see some residual cosmology dependence. The latter depends on $B_{YX}$ as well. For instance, the \LMgas\ relation shows $B_{LM\text{gas}}\sim 1.2$ (e.g., \cite{zhang11,lovisari15}), which results in a $\sim \log{H_0}$ dependence. One can apply the same exercise to any scaling relation. 

It is evident from the above that, using just one galaxy cluster sample with multiple measured quantities per cluster, one can obtain complementary information on cosmic isotropy. The different isotropy constraints also yield excellent consistency checks and reveal possible systematic and cluster physics biases, further reinforcing the robustness of the methodology. The plethora of (nearly independent) cosmological information given per object constitutes a substantial advantage of clusters as probes of cosmic isotropy over other extra-galactic objects.

\subsubsection{$H_0$ anisotropies}\label{H0_anis}
The dominant cosmological effect in Eq. \ref{LT-anisotropy} and other similar scaling relations comes from the $N_{YX}\log{H_0}$ term. $N_{YX}$ varies according to the used scaling relation, as explained before. Absolute constraints can be put on the sum $N_{YX}\log{H_0}+A_{YX}$ but not on $H_0$ alone without assuming the value of $A_{YX}$. However, $A_{YX}$ is not expected to vary with direction. $A_{YX}$ depends on inner-cluster (mostly baryonic) physics, such as cluster gas thermodynamics, and the mechanisms that connect the scaling cluster properties. Thus, the true, "intrinsic" $A_{YX}$ is independent of the spacetime between the cluster and the observer. As a result, $A_{YX}$ carries no information about the redshift-distance relation. Assuming that clusters are locally described by the same physical laws, $A_{YX}$ should not have any directional dependence, i.e., it should be statistically isotropic. Thus, when statistically significant angular variations of $N_{YX}\log{H_0}+A_{YX}$ are observed, the bulk of the effect can be attributed to an $H_0$ anisotropy. In practice, one keeps $A_{YX}$ fixed and treats $H_0$ as a free parameter across the sky. As explained above, the absolute $H_0$ values are arbitrary and depend on the exact choice of $A_{YX}$; for example, $A_{YX}$ can be chosen so the full-sky cluster sample returns $H_0=70$ km/s/Mpc. On the other hand, relative angular $H_0$ variations are directly constrained and are completely independent of the assumed $A_{YX}$ value.

Finally, an angular variation of $H_0$ is degenerate with bulk flows at $z\lesssim 0.1$ scales and almost entirely independent of bulk flows at $z\gtrsim 0.2$ (see Sect. \ref{bulk flows}). On the other hand, there is strong degeneracy between $H_0$ and $\Omega_{\text{m}}$ anisotropies at $z\gtrsim 0.4$ while there is almost none at $z\lesssim 0.2$ (see Sect. \ref{Omega}). Thus, at $z\approx 0.2-0.3$, one can purely trace $H_0$ anisotropies with no strong effects from other possible cosmological anisotropies. At more local or distant scales, one can independently trace anisotropies on the redshift-distance relation as a whole. However, to attribute such an apparent anisotropy to an individual cosmological parameter (or bulk flow), one needs to assume the other degenerate cosmological parameter to be fixed or vary within a limited range.

\subsubsection{Bulk flows}\label{bulk flows}
Bulk flows can have a substantial impact on the directional behavior of $G(\Omega_{\text{m}},C_{YX},z)$ and scaling relations. Bulk flows can affect the inference of $z$ due to large (unaccounted) cluster peculiar velocities with non-negligible $z_{\text{pec}}$ compared to $z$ (Eq. \ref{redshifts}). Such effects are mostly relevant for local cluster samples ($z\lesssim 0.15$). For higher$-z$ clusters peculiar velocities become irrelevant compared to the Hubble recession velocity. In essence, even if strong bulk flows exist at $z\gtrsim 0.2$, they will not cause any detectable anisotropy to scaling relations.

A biased $z$ due to bulk flow presence strongly affects the inferred cluster distance through the integration limits of $\int_0^{z} \frac{dz'}{E(z')}$. At the same time, it leaves $E(z)$ nearly unaffected, constituting the exact redshift evolution term $C_{YX}\log{E(z)}$ insignificant for bulk flow studies. As an example, a $+1000$ km/s bulk flow at $z=0.05$ would cause a $\approx 7\%$ overestimation of $\int_0^{z} \frac{dz'}{E(z')}$ and only a $<0.2\%$ change in $E(z)$. As previously mentioned, bulk flows at $z\lesssim 0.2$ scales have similar effects on scaling relations as an $H_0$ angular variation. For example, a $7\%$ $H_0$ anisotropy would also cause a $7\%$ change in the inferred cluster distance, having the same effect with a $+1000$ km/s bulk flow at $z=0.05$. Consequently, one cannot easily distinguish between the two phenomena at local scales. Nevertheless, when more distant clusters are studied, the effect of bulk flows fades out while an $H_0$ anisotropy has the same effect for every cosmic scale. Characteristically, the same $+1000$ km/s bulk flow would only cause a $1.6\%$ change in $D_L$ at $z=0.2$.

\subsubsection{$\Omega_{\text{m}}$, $\Omega_{\Lambda}$, and $w$ anisotropies}\label{Omega}

An angular variation of the $E(z)$ factor (i.e., $\Omega_{\text{m}}$, $\Omega_{\Lambda}$, and the dark energy equation-of-state $w$) will cause an anisotropic effect on $G(\Omega_{\text{m}},C_{YX},z)$ under certain conditions. For local cluster samples ($z\lesssim 0.3$), $E(z)$ variations only weakly affect the directional behaviour of most scaling relations, usually on a lower level than statistical noise. At higher$-z$, directional variations in $E(z)$ become more impacting on $G(\Omega_{\text{m}},C_{YX},z)$ and can be detected above the statistical noise level. $E(z)$ affects $G(\Omega_{\text{m}},C_{YX},z)$ through both $\log{\left[\int_0^{z} \frac{dz'}{E(z')}\right]}$ and $C_{YX}\log{E(z)}$. However, for scaling relations with $C_{YX}>0$ (e.g., \YT\ and \MgasT), changes in $E(z)$ tend to partially cancel out between the two terms, suppressing the impact of an $E(z)$ anisotropy on scaling relations even at high$-z$. On the other hand, for scaling relations with $C_{YX}<0$ (e.g., \LT), the impact of $E(z)$ on $G(\Omega_{\text{m}},C_{YX},z)$ is enhanced. Consequently, by studying the anisotropic behaviour of $C_{YX}<0$ and $C_{YX}>0$ scaling relations, it is trivial to conclusively distinguish between an $E(z)$ anisotropy and a bulk flow or an $H_0$ anisotropy (for which $C_{YX}$ is irrelevant). The value of $C_{YX}$ can either assumed to be known from the self-similar model \cite{kaiser}, adopted from a past study that constrained it (e.g., \cite{reichert}), or treat it as a free parameter. Nonetheless, as mentioned before, the effects of $E(z)$ anisotropies remain small for $z\lesssim 0.3$ samples for any $C_{YX}$. For now, there is very limited availability of adequately large, all-sky, high$-z$ cluster samples with which $E(z)$ anisotropies could be studied. 

Changes in $\Omega_{\text{m}}$ could also cause a directional dependence of the true $A_{YX}$ and $B_{YX}$ values for scaling relations between properties that trace the total amount of gas (e.g., \Lx) and the total cluster mass (e.g., $M_{\text{tot}}$ or even \Tx), due to the different gas fraction per direction. These additional effects might be challenging to take into account. For all these reasons, the cosmological parameters included in $E(z)$ are often assumed to be fixed and any observed anisotropies are attributed to $H_0$ variations or bulk flows which have a much stronger influence at $z\lesssim 0.3$ clusters anyway.

\subsubsection{When cluster physics effects and systematic biases dominate}\label{systematic_anisotropies}

There are scaling relations that exhibit negligible cosmological dependence and can be utilised to detect systematic biases or cluster physics effects. These relations are particularly helpful for disentangling the signal from the possible cosmological anisotropies from other cluster-related causes that might influence the conclusions on isotropy. For example, the \LY\ relation has $B_{LY}\sim 0.93$ \cite{migkas21} which results in a $\lesssim 0.1\log{H_0}$ dependency. Consequently, the \LY\ relation is insensitive to cosmological anisotropies. However, it can effectively trace biases in the X-ray Galactic absorption correction applied to \Lx. While the latter is subject to Galactic absorption effects, \Ysz\ is not since it is measured in microwaves. Thus, a statistically significant angular variation of \LY\ can provide the true X-ray absorption per direction by revealing a systematic over- or underestimation of the relative correction factor for different sky patches. Similarly, the \RMgas\ relation yields a rather weak $0.4\log{H_0}$ dependency (Migkas et al. in prep.). Hence, its angular variation level is mainly dictated by the average morphology of the cluster population per direction. Cluster morphology (e.g., relaxed or disturbed and cool core or non-cool core) strongly affects $R_{50\%}$ while it does not cause considerable changes to $M_{\text{gas}}$. Therefore, it is essential to robustly check if the cosmological sensitivity of a scaling relation is higher than its sensitivity on different cluster physics effects and systematics before we conduct isotropy tests.

\subsection{Properties of cluster samples used for isotropy tests}
To carry out an unbiased test of cosmic isotropy using galaxy cluster samples, the latter should ideally fulfil certain criteria. Firstly, a similar cluster population across the sky is needed. This ensures that $A_{YX}$ is similar in every direction. A homogeneous cluster selection, independent of direction, generally achieves such a cluster population uniformity. It also provides similar numbers of clusters in every direction, allowing for balanced full-sky coverage, which is a necessity for an accurate isotropy test. Even if a sample is not uniformly built, one can apply selection cuts to obtain a uniform cluster population across different sky patches before conducting isotropy tests.

If a cluster sample is constructed with an inhomogeneous selection, the sample might suffer from a directional dependence on cluster properties. For instance, more relaxed/disturbed clusters or more intrinsically brighter/fainter clusters may be found in a specific direction than the rest of the sky. Such an imbalance in the average cluster properties will cause an anisotropic $A_{YX}$ since distinct cluster populations are described by somewhat different scaling relations. If one ignores the spatial variation of $A_{YX}$, the inferred cosmological anisotropy will be overestimated. However, even in a homogeneously selected cluster sample, $A_{YX}$ can slightly vary across the sky due to the random variations of the cluster population. That is, $A_{YX}$ can slightly vary across the sky due to statistical noise. This effect is small for low-scattered scaling relations, but it should (and can) be taken into account when inferring the angular variation of cosmological parameters. If it is assumed to be negligible, it might lead to an overestimation of the apparent cosmological anisotropies.

Another useful feature of cluster samples used for cosmic isotropy studies is the similarity of the $z$ distribution across the sky. This serves two purposes. Firstly, if an inaccurate $C_{YX}$ value is adopted and a rather high$-z$ sample is studied, a dissimilar $z$ distribution will induce directional biases in the scaling relation parameters. These could be wrongfully interpreted as cosmological anisotropies. On the contrary, if the $z$ distribution is consistent across the sky, a wrong $C_{YX}$ value will cause the same bias towards every direction, which will eventually cancel out when comparing independent sky regions; that is, the cosmological anisotropy constraints will remain unaffected. Nevertheless, for local cluster samples ($z\lesssim 0.2$), $C_{YX}$ biases become insignificant since no strong cluster evolution is expected in such a local cosmic volume. The second reason why a similar $z$ is essential is that, when comparing cosmological parameters across different sky patches, one wishes to trace similar cosmic volumes. Let us assume a hypothetical anisotropy is present only at, e.g., $z\lesssim 0.1$ due to bulk flows. Then, if clusters in one direction lie at greater distances on average, they will not be affected by such a bulk flow. On the other hand, if more local clusters are found in a different direction, they will be affected by the bulk flow. This will bias the detection of such a cosmological effect.

\section{Evidence for local anisotropy from galaxy clusters}

\subsection{First application and preliminary findings}
Probing cosmic isotropy with galaxy cluster scaling relations is a relatively new sub-field of cosmology. The idea was introduced in \cite{migkas18} where they used the \LT\ relation to test cosmic isotropy with two independent cluster samples; the ASCA Cluster Catalog (ACC, \cite{horner}) and the first data release of the XMM Cluster Survey sample (XCS, \cite{XCS}). ACC is an archival sample containing 272 clusters with a rather uniform sky distribution and median $z\approx 0.09$. XCS was built from archival \textit{XMM-Newton} data and contains 364 with \Lx\ and \Tx\ measurements. These clusters cover the full sky but not uniformly, with some regions having many more objects than others. XCS is a rather high$-z$ sample with median $z\approx 0.35$. \cite{migkas18} performed a 1D analysis with respect to the Galactic longitude $l$ measuring $H_0$ and $\Omega_{\text{m}}$ across different $l$ bins. Interestingly, they identified the same anisotropy patterns in both cluster samples despite the latter being independent while tracing different cosmic scales. They found $H_0$ to vary by $\sim 20\%$ at a mild $\approx 2.5-3\sigma$ level, with the discrepancy being more prominent at $z\lesssim 0.35$. This analysis offered a first glance at the potential of cluster scaling relations as proxies of cosmic isotropy; nonetheless, the results were preliminary since both used samples suffered from certain weaknesses. For example, ACC was compiled by old, low-quality X-ray cluster measurements with no specific selection. Also, \Tx\ was measured within a non-physical cluster radius. On the other hand, XCS composed of low-mass galaxy clusters and groups that generally suffer from larger scaling relation dispersion. Also, the \Tx\ was measured with large uncertainty in a radius that strongly depended on the fiducial cosmology (within 300 kpc from the cluster's center); several clusters also had photometric $z$ instead of spectroscopic. Finally, both samples did not have sufficient information to robustly assess the cluster population characteristics per direction. Thus, there was still much room for progressing this methodology.

\subsection{First 2D results and improved cluster catalogues}
Results on apparent $H_0$ variations from the \LT\ scaling relation were significantly improved by \cite{migkas20}. That study presented and utilised an X-ray flux-limited, homogeneously-selected local cluster sample, namely, a preliminary version of the extremely expanded HIgh FLUx Galaxy Cluster Sample (eeHIFLUGCSv0). The latter consists of 313 clusters with self-consistently measured, high-quality \Tx\ measurements with low statistical uncertainties as obtained using deep \textit{XMM-Newton} and \textit{Chandra} data. The 313 clusters cover the full sky uniformly and have a median $z\approx 0.075$ while almost all clusters lie at $z\lesssim 0.3$. The \LT\ scatter was shown to be much lower for eeHIFLUGCSv0 than ACC and XCS. Adequate information to characterise the cluster population per direction was also provided. Moreover, \cite{migkas20} implemented cones of varying angular sizes covering all directions in the sky, providing the first 2D results on apparent $H_0$ variations from cluster scaling relations. For eeHIFLUGCSv0, they identified a $\approx 13\%$ $H_0$ maximum angular variation with a statistical significance of $3.59\sigma$ towards $(l,b)\sim (281^{\circ}, -16^{\circ})$. This finding was consistent with the directions identified by \cite{migkas18}. The anisotropy was also present when $z<0.03$ clusters were excluded, suggesting that the bulk of the effect continues beyond $\gtrsim 130$ Mpc. Jointly analysing the 842 independent cluster samples from eeHIFLUGCSv0, ACC, and XCS, \cite{migkas20} found a combined $\approx 14\%$ apparent $H_0$ variation towards $(l,b)\sim (312^{\circ}, -21^{\circ})$ at a $4.55\sigma$ level. However, this finding did not assume the shape of the anisotropy as they simply compared the two most extreme, independent sky regions. When a dipole anisotropy was assumed, the $H_0$ variation and statistical significance dropped to $\approx 11\%$ $4\sigma$ level, respectively. \cite{migkas20} explored an extensive list of possible systematic biases but failed to identify any bias that would alleviate the tension with the isotropy hypothesis. Overall, this work confirmed the strong indications for an anisotropy in the late Universe, as traced by galaxy clusters. Nevertheless, further confirmation was needed and remaining systematics were still to be addressed.

\subsection{More precise results from multi-wavelength cluster scaling relations}
\begin{figure}[h]
\centering \includegraphics[width=3.7in]{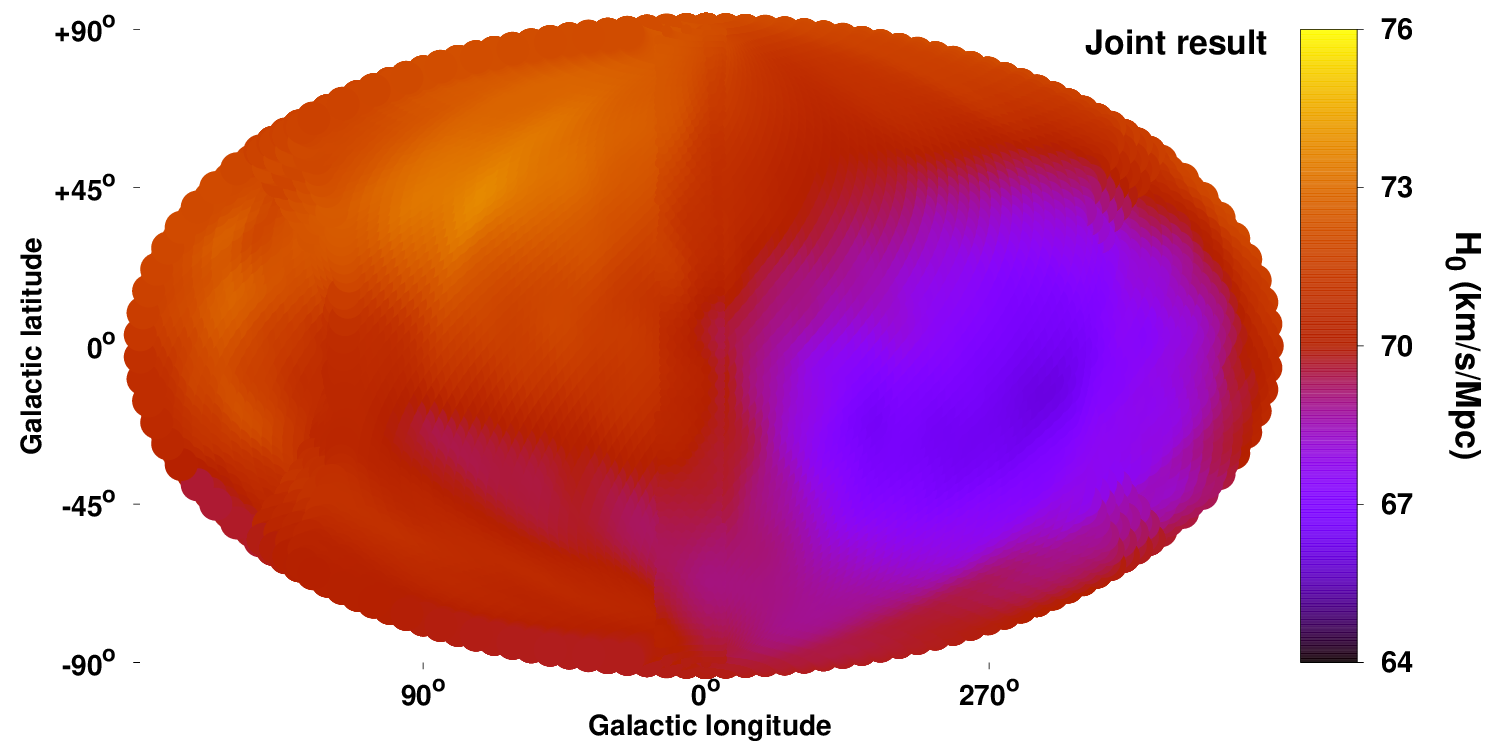}
\caption{$H_0$ anisotropy map as derived by \cite{migkas21} from the joint analysis of different multiwavelength scaling relations and cluster samples. A $9\%$ $H_0$ dipole was detected with a statistical significance of $5.4\sigma$ (credit: Migkas et al., 2021, A\&A, 649, A151).}
\label{H0-plot-M21}
\end{figure}

The most robust evidence for a local Universe anisotropy using galaxy clusters was presented by \cite{migkas21}. In that study, they used the eeHIFLUGCSv0 sample as presented in \cite{migkas20} and additionally measured the microwave \Ysz, infrared \Lbcg, and X-ray $R_{50\%}$ for each cluster using \textit{Planck}, 2MASS, and \textit{ROSAT} data respectively. They also measured \Ysz\ for the ACC sample. These new measurements allowed the construction of ten multi-wavelength scaling relations. The statistical methodology was also noticeably improved compared to past studies. Combining all the available information by properly accounting for all scatter and measurement covariance, \cite{migkas21} detected an apparent $9\%\ H_0$ dipole anisotropy towards $(l,b)\sim (280^{\circ}, -15^{\circ})$ at an unprecedented $5.4\sigma$ level (displayed in Fig. \ref{H0-plot-M21}). They assessed the statistical significance of the $H_0$ dipole by employing isotropic Monte Carlo simulations. Furthermore, \cite{migkas21} fully exploited 14 cosmology-independent cluster properties (e.g., relaxation state, Galactic absorption, metallicity, exposure time, and cluster core \Tx) to predict the expected $H_0$ behaviour of a cluster sub-sample purely based on cluster population criteria. This sophisticated approach confirmed that the emerging $H_0$ dipole is not significantly affected by cluster population differences across the sky. Overall, numerous tests were performed to ensure the validity of their results.

\begin{figure}[h]
\centering \includegraphics[width=2.6in]{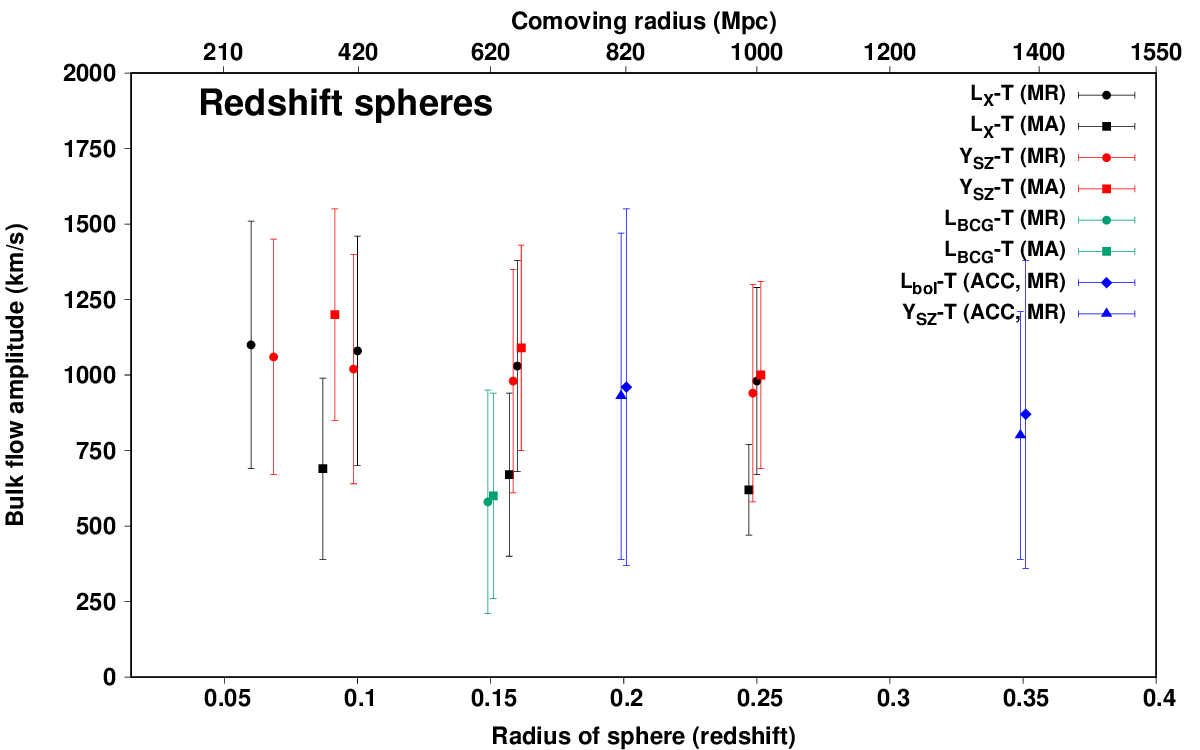}
\includegraphics[width=2.6in]{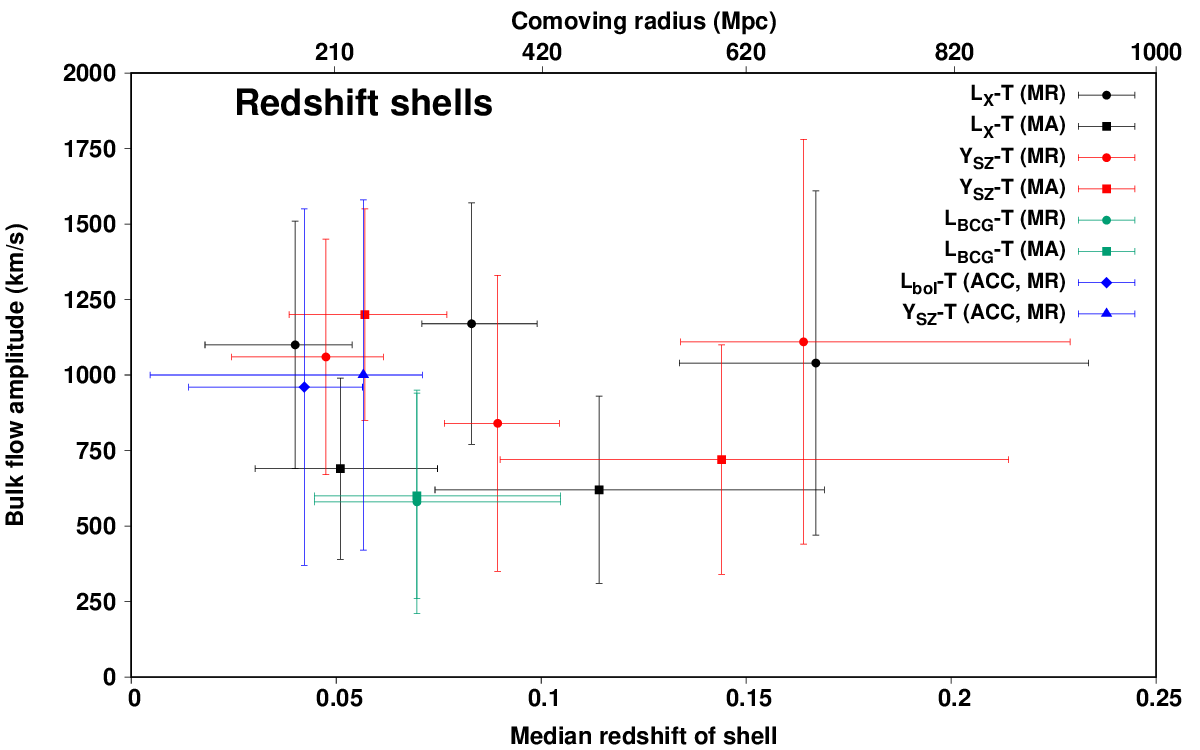}
\caption{\textit{Left}: Bulk flow amplitude and its $1\sigma$ uncertainty as a function of the redshift (or comoving distance) radius as determined by \cite{migkas21}. The results from different scaling relations, cluster samples, and statistical approaches are displayed in different shapes and colours. 
\textit{Right}: Same as in left panel but as a function of redshift (or comoving distance) shells instead of spheres (credit: Migkas et al., 2021, A\&A, 649, A151).}
\label{BF-plot-M21}
\end{figure}

Since eeHIFLUGCSv0 is a local cluster sample, one can attribute the apparent $H_0$ anisotropy to large cluster bulk flow. As explained in Sect. \ref{bulk flows}, the two cosmological effects are nearly indistinguishable at $z\lesssim 0.2$. \cite{migkas21} explored this scenario and provided the first bulk flow constraints from galaxy cluster scaling relations. They found that the apparent $H_0$ dipole could emerge due to a $\approx 900\pm 200$ km/s bulk flow extending beyond 500 Mpc ($z\gtrsim 0.12$). As displayed in Fig. \ref{BF-plot-M21}, they constrained the cluster bulk flow as a function of different cosmic volumes and shells. Different scaling relations and independent cluster samples returned consistent results with the best-fit bulk flow amplitude ranging from $\sim 600-1200$ km/s. They found that the bulk flow direction did not significantly deviate within $z\sim 0.05-0.15$ scales. However, they noted that the results beyond $z\gtrsim 0.1$ are likely overestimated as they are dominated by lower$-z$ objects. However, there is no available comparison with another independent study yet since galaxy clusters are the only probes for now that cover these cosmic scales uniformly across the sky. Overall, this study provided the strongest evidence until now that the cluster rest frame does not coincide with the CMB rest frame at $z\lesssim 0.15$, which is in strong contrast with $\Lambda$CDM.

It is important to note that the $5.4\sigma$ statistical significance of the apparent $H_0$ dipole detection in \cite{migkas21} does not directly correspond to the tension with $\Lambda$CDM. The latter still allows for limited residual anisotropy due to bulk flow motions at these scales. For $\sim 500$ Mpc volumes ($z\lesssim 0.12$), $\Lambda$CDM predicts an average $\approx 80$ km/s bulk flow which roughly corresponds to a $\sim 1\%$ dipole $H_0$ anisotropy. Consequently, the findings of \cite{migkas21} are at a $\sim 4.5\sigma$ tension with $\Lambda$CDM. Finally, they fixed $A_{YX}$ for all scaling relations; this might underestimate the $H_0$ dipole uncertainty. However, the final statistical significance is not expected to be overestimated since they follow the same procedure using isotropic Monte Carlo simulations. 

To conclude, past studies using various multi-wavelength scaling relations and independent cluster samples strongly and consistently indicate an apparent $H_0$ angular variation within $\lesssim 1$ Gpc. However, the apparent anisotropy can be alleviated by a considerable cluster bulk flow motion, although the latter is in notable tension with $\Lambda$CDM as well. Future studies with new data and further improved methodologies will reveal if the observed dipole anisotropy truly constitutes a remarkable failure of the standard cosmological model in the local Universe or if its statistical significance has been overestimated.

\section{Future Outlook}
We live in the era of next-generation, state-of-the-art cosmological surveys that remarkably increase the amount of available data and aim at precise cosmological constraints. Furthermore, recent hydrodynamical cosmological simulations offer a unique tool to help us better understand observational findings and uncover potential systematic biases. As a result, there is vast potential for scrutinising cosmic isotropy with galaxy clusters using various approaches. Moreover, new advances in the statistical and modelling methodology might offer new valuable insights. Here we present a non-exhaustive list of expected future advances in this sub-field that will yield significant results within the next years.

\subsection{New cluster data from multi-wavelength surveys}\label{future_surveys}
There are numerous cosmological experiments that will soon provide (or have already done so) valuable new cluster data. First and foremost, the extended ROentgen Survey with an Imaging Telescope Array (\textit{eROSITA}, \cite{predehl21}) and its planned 4-year \textit{eROSITA} All-Sky Survey (eRASS) will substantially progress cluster science owing to the largest X-ray cluster catalogues ever compiled. Already from the first all-sky scan (eRASS1), \textit{eROSITA} has detected $>12,000$ X-ray clusters, with $>8,000$ of them as new detections \cite{bulbul}. Almost all of these clusters have \Lx\ and $M_{\text{gas}}$ measurements, while for $\sim 450$ of them, precise \Tx\ measurements are already available.\footnote{\textit{eROSITA} \Tx\ measurements were recently found to deviate from \Tx\ measured by other major X-ray telescopes, such as \textit{Chandra} and \textit{XMM-Newton} \cite{migkas24}. However, this deviation is accurately known and one can correct for it, not jeopardising isotropy tests with \textit{eROSITA} \Tx\ measurements.} These numbers will significantly increase as deeper eRASS data are analysed, reaching $\gtrsim 50,000$ massive clusters for eRASS:8 \cite{pillepich}. Additionally, $1499$ eRASS1 clusters have $\sigma_\text{vel}$ measurements \cite{kluge} that will allow the study of X-ray-optical scaling relations, such as $L_{\text{X}}-\sigma_\text{vel}$. Due to the large amount of new X-ray cluster information \textit{eROSITA} will return, a significant improvement in cosmic isotropy tests with X-ray clusters is expected.

In the optical and infrared regime, the \textit{Euclid} telescope and its \textit{Euclid} Wide Survey will scan $\approx 15,000^{\circ}$ in the extra-galactic sky, discovering $\gtrsim 10^5$ massive clusters ($M_{\text{tot}}\gtrsim 2\times 10^{14}\ M_{\odot}$) at $z>0.2$ \cite{sartoris}. Additionally, $\gtrsim 10^4$ of these clusters will be at $z>1.3$. For a large fraction of them, \textit{Euclid} will measure $\sigma_{\text{vel}}$ and $\lambda$ values, as well as $M_{\text{tot}}$ through weak lensing data. This will allow for constructing optical cluster scaling relations with vast numbers of used objects. As a result, \textit{Euclid} will allow for an unprecedented characterisation of isotropy at large cosmic scales, insensitive to local effects, such as bulk flows. Other optical surveys that (will) cover large areas of the sky, such as the ones from \textit{Rubin} Observatory and \textit{DESI}, will also detect tens and hundreds of thousands of clusters, respectively, measuring useful optical properties for many of them \cite{LSST,zou,wen}. It is evident that the abundance of optical galaxy cluster information to become available in the next years is striking and will immensely advance our understanding of cosmic isotropy at large scales.

In microwave, the Atacama Cosmology Telescope (\textit{ACT}) has already provided $>4000$ SZ-detected clusters with \Ysz\ measurements and a median $z\approx 0.5$ \cite{hilton}. \textit{ACT}significantly increased the availability of \Ysz\ values compared to past cluster samples. The efficiency of X-ray-SZ scaling relations for testing cosmic isotropy was demonstrated in \cite{migkas21}. Finally, this list is non-exhaustive since many more cosmological experiments will provide valuable cluster data in the near future.

\subsection{Low-scattered scaling relations and improved modelling}
There are two primary approaches to improve constraints on cosmic isotropy; firstly, employ new, larger samples as outlined in Sect. \ref{future_surveys}, and secondly, reduce the scaling relation scatter of current samples. The constraining power roughly scales with $\sim \delta/\sqrt{N}$, where $\delta$ is a relation's dispersion (or scatter) and $N$ is the number of used objects. Hence, lowering the scatter of a scaling relation to half has the same constraining effect as increasing the sample size by a factor of four. Consequently, it is evident that suppressing the scatter of current scaling relations with improved modelling is highly important. 

Certain scaling relations with lower scatter than those utilised in past studies are expected to return valuable insights in the near future. For example, the \MgasT\ scaling relation is particularly useful for cosmic isotropy tests due to its low $H_0$ dispersion ($\delta_{H0}\approx 11\%$). Moreover, the \LT\ relation (and other X-ray relations) exhibits significantly reduced scatter when the core-excised \Lx\ is used instead of the total \Lx\ \cite{pratt}. Hence, core-excised cluster measurements can be implemented to improve existing isotropy results. Furthermore, cluster morphology parameters, such as the X-ray concentration and centroid shift, correlate with the scatter of some scaling relations \cite{bharad,zhu,yuan}. A similar scatter correlation has been found with the cluster dynamical state \cite{zhang,lovisari20,migkas21,poon}. If one measures the cluster morphology and includes it in the scaling relation model, this could potentially reduce the scatter and enhance the cosmological constraints. 


\subsection{Cosmological simulations}
The validity of methodologies that search for apparent, statistically significant cosmic anisotropies should be cross-checked using cosmological simulations. Recently, the state-of-the-art hydrodynamical FLAMINGO simulations were introduced \cite{schaye}. FLAMINGO recreate different realisations of isotropic $\Lambda$CDM universes with $\sim 10^9\ M_{\odot}$ mass resolution and implements full baryonic physics effects. The simulations also provide lightcones for multiple different observers reaching out to 2.8 Gpc ($z\sim 0.8$) and for different cosmologies. Numerous properties for each simulated halo are available for each lightcone, including all the cluster properties discussed in this review. As a result, different methods for detecting cosmic anisotropies can be applied to the FLAMINGO data to investigate the robustness of each method. Such a test will also assess the true rarity of the observed anisotropies within a $\Lambda$CDM universe. 

\section{Discussion and Summary}
Galaxy clusters and their scaling relations offer an extraordinary tool for probing cosmic isotropy at a wide range of cosmic scales. For now, galaxy clusters are the only isotropy probe for which we have uniform, all-sky data at $z\sim 0.05-0.5$. Moreover, clusters have a unique advantage over other isotropy probes; they can provide information on the angular variation of cosmological parameters across many different wavelengths. This enables the construction of numerous multi-wavelength scaling relations that probe cosmic isotropy almost independently, with only a single cluster sample. Low-scattered scaling relations between cosmology-dependent (e.g., \Lx, \Ysz, and $M_{\text{gas}}$) and cosmology-independent (e.g., \Tx\ and $\sigma_{\text{vel}}$) cluster properties are exceptionally useful for such tests. Nonetheless, every scaling relation can be decomposed into observables and cosmological parameters. Thus, by determining the angular variation of each scaling relation, robust conclusions on the anisotropy of cosmological parameters and the existence of large-scale bulk flows can be drawn. Angular $H_0$ variations identically affect all cosmic scales, while bulk flows and $\Omega_{\text{m}}$ anisotropies mainly affect $z\lesssim 0.2$ and $z\gtrsim 0.3$ scales, respectively. Scaling relations that are insensitive to cosmological anisotropies (e.g., \LY\ and \RMgas) can reveal multiple directional cluster physics effects and help us isolate the cosmological signal from such potential biases.

There are, however, certain pitfalls in this method. The most critical is that the used sample should show a similar cluster population throughout the sky. Varying cluster populations (e.g., relaxed or disturbed, low or high mass, and centrally peaked or diffuse) are often (but not always) described by slightly different scaling relations. A uniform cluster selection that results in similar cluster populations is essential to ensure that the underlying scaling relations of the independent cluster sub-samples are statistically equivalent. Thus, any observed difference can be then attributed to cosmological effects. If there are variations in the cluster population between compared regions, the inference of any cosmological effect can be distorted. Correspondingly, compared cluster sub-samples must also have comparable $z$ distributions to make certain we trace the same cosmic scales per direction.

Recent studies used several multi-wavelength cluster scaling relations to probe the isotropy of the local ($z\lesssim 0.3$) Universe. Specifically, \cite{migkas21} reported the detection of a $9.0\pm 1.7\%$ $H_0$ dipole using X-ray, SZ, and infrared cluster data, substantially challenging the convergence to isotropy hypothesis at these cosmic scales. They argued that the apparent $H_0$ variation might be caused by a $\sim 900$ km/s cluster bulk flow extending to $500$ Mpc, at $\gtrsim 4\sigma$ tension with $\Lambda$CDM.

New cluster samples of unprecedented sizes from current and future surveys will considerably boost the cosmic isotropy constraints from scaling relations. Such surveys include \textit{eROSITA} and \textit{Euclid} that will eventually measure X-ray and optical/infrared cluster quantities for $\gtrsim 50,000$ and $\gtrsim 10^5$ clusters, respectively. Moreover, improvements in scaling relations can be achieved by considering the cluster morphology and dynamical state. Such refined modelling will result in lower scatter and more precise cosmological constraints.

Finally, utilising large, cosmological, hydrodynamical simulations will enable testing the robustness of different anisotropy detection methods using cluster scaling relations. This is crucial in order to understand the importance of the observational findings, eliminating possible systematic biases in the applied methodology. Overall, multi-wavelength galaxy cluster scaling relations will play a pivotal role in testing the isotropy of the late Universe in the near future.

\vskip6pt

\ack{I acknowledge support in the
form of the X-ray Oort Fellowship at Leiden Observatory. I would also like to thank the Royal Society and the organisers for inviting me as a speaker to the "Challenging the standard cosmological model" meeting, which sparked the writing of this review. Finally, I thank my wife for taking several vacation days off her job to care for our son so I can finish writing this review on time.}


\bibliographystyle{RS}
\bibliography{sample} 






\end{document}